\documentclass[useAMS,usegraphicx]{mn2e}
\usepackage{amssymb}
\newcommand{\mnras}{MNRAS}
\newcommand{\apj}{ApJ}
\newcommand{\aaps}{A\&AS}

\newcommand{\aj}{AJ}
\newcommand{\apjl}{ApJ}
\newcommand{\apjs}{ApJS}

\newcommand{\sex}{SEXTRACTOR\ }
\newcommand{\gal}{GALFIT\ }

\title[PyMorph]{PyMorph: Automated Galaxy Structural Parameter Estimation using
Python}
\author[Vinu Vikram et al.]{Vinu Vikram,$^{1,3}$\thanks{E-mail:
vvinuv@iucaa.ernet.in} 
Yogesh Wadadekar,$^{2}$\thanks{E-mail: yogesh@ncra.tifr.res.in} 
Ajit K. Kembhavi$^{3}$\thanks{E-mail: akk@iucaa.ernet.in} 
\newauthor
and G. V. Vijayagovindan$^{1}$\thanks{Deceased}\\
$^{1}$School of Pure and Applied Physics, Mahatma Gandhi University, Kottayam
686560, India\\
$^{2}$National Centre for Radio Astrophysics, Post Bag 3, Ganeshkhind, Pune
411007, India\\
$^{3}$Inter University Centre for Astronomy and Astrophysics, 
Post Bag 4, Ganeshkhind, Pune 411007, India}

\begin{document}
\date{Accepted 2010 July 23.  Received 2010 July 20; in original form 2010 April 6}

%\pagerange{\pageref{firstpage}--\pageref{lastpage}} \pubyear{2009}

\maketitle

\label{firstpage}

\begin{abstract}

We present a new software pipeline -- PyMorph -- for automated
estimation of structural parameters of galaxies. Both parametric fits
through a two dimensional bulge disk decomposition as well as
structural parameter measurements like concentration, asymmetry
etc. are supported. The pipeline is designed to be easy to use yet
flexible; individual software modules can be replaced with ease. A
find-and-fit mode is available so that all galaxies in a image can be
measured with a simple command. A parallel version of the Pymorph
pipeline runs on computer clusters and a Virtual Observatory
compatible web enabled interface is under development.
   
\end{abstract}


\section{Introduction}

In recent years, the morphological analysis of galaxies has provided
invaluable information regarding their origin and evolution. This analysis 
used large galaxy samples drawn from modern
astronomical surveys such as the SDSS \nocite{yor00, aba09}({York} {et~al.} 2000; {Abazajian} {et~al.} 2009), 2MASS
\nocite{skr06}({Skrutskie} {et~al.} 2006), GEMS \nocite{rix04}({Rix} {et~al.} 2004), COSMOS \nocite{sco07}({Scoville} {et~al.} 2007) etc. Since
visual estimation of galaxy morphology is most accurate, it is widely
used. But given the large numbers of galaxies in astronomical
datasets, it is impractical to classify them all using human
classifiers (unless a large volunteer base is available, like in the
GalaxyZoo project \nocite{lin08}({Lintott} {et~al.} 2008)). Besides, different human classifiers may not
agree
completely on the morphological classification. It is therefore
desirable to develop a reliable, objective and automated method for
quantitative morphological classification.

It has been known for a long time that the visual morphology of
galaxies is well correlated with their physical properties. For
example, the colour correlates with the morphology such that late type
spirals are bluer, on average, than elliptical galaxies. Similarly, we
can make use of the surface brightness profile of galaxies to classify
them.

In general, the stellar component of galaxies can be decomposed into a
bulge and a disk. While the disk profile is usually an exponential,
the bulge is well approximated by the S\'{e}rsic function
\nocite{ser68}({Sersic} 1968).  It is found that the bulge-to-total luminosity ratio
($B/T$) of galaxies correlates with the visual morphology where the
$B$ is the light contained in the bulge component and $T$ is the total
light of the galaxy. Elliptical galaxies are expected to have $B/T
\simeq 1$, while pure disk galaxies have $B/T \simeq 0$.  In recent
years, most researchers prefer to fit a two dimensional representation
of the bulge and disk profiles directly to a broad band image of the
galaxy \nocite{byu95,jon96,wad99,pen02,sou04}({Byun} \& {Freeman} 1995; {de Jong} 1996; {Wadadekar}, {Robbason} \&  {Kembhavi} 1999; {Peng} {et~al.} 2002; {de Souza}, {Gadotti}, \& {dos  Anjos} 2004). 
This structural decomposition technique is not only useful
for quantifying the morphology but is also an excellent tool for
studying the formation and evolution of galaxies of different
morphological types \nocite{kho00,rav01,sim02,mac03,bar07,bar09,vik10}(eg. {Khosroshahi}, {Wadadekar} \&  {Kembhavi} 2000; {Ravindranath} {et~al.} 2001; {Simard} {et~al.} 2002; {MacArthur}, {Courteau}, \&  {Holtzman} 2003; {Barway} {et~al.} 2007, 2009; {Vikram} {et~al.} 2010).
Recent research has shown that the simple  
S\'{e}rsic bulge + exponential disk formulation is not adequate for many
galaxies \nocite{lau05,gad08,pen10}({Laurikainen}, {Salo} \& {Buta} 2005; {Gadotti} 2008; {Peng} {et~al.} 2010) and fitting these
simple models can lead to wrong estimates of structural parameters.
Nevertheless, these simple models, when used appropriately, do give reliable
results and are useful indicators of galaxy structure.

Since parametric methods (such as the two dimensional bulge disk
decomposition) are generally computational intensive and difficult to
apply to small faint galaxies, several non-parametric methods have
been developed during the last few years to quantify galaxy
morphology. The main motivation for the development of these
non-parametric methods is to make classification possible at very high
redshifts where the images do not have enough resolution elements and
signal-to-noise for visual classification
\nocite{abr96,con03,lot04}({Abraham} {et~al.} 1996; {Conselice} 2003; {Lotz}, {Primack} \& {Madau} 2004). Non-parametric methods are not
computationally intensive compared to the parametric methods. However,
with non-parametric methods, it is not easy to convert measured
quantities to physically meaningful parameters such as bulge or disk
luminosity.

In this paper, we describe an automated pipeline software PyMorph to
estimate structural parameters of galaxies. We have developed this
pipeline by glueing together widely used codes such as \sex
\nocite{ber96}({Bertin} \& {Arnouts} 1996) and \gal \nocite{pen02}({Peng} {et~al.} 2002)\footnote{Although PyMorph currently uses
version 2.5.0 of \sex and version 2.03b of \gal, it can easily be modified to
use newer versions of these codes.} to our own codes for automation
and quality control. We have also developed our own implementation of
non-parametric methods (Section \ref{nonparam}). In Section
\ref{param}, we explain the operational procedure to obtain structural
parameters using \gal and SEXTRACTOR. In Section~\ref{setup}, we describe
how to setup the pipeline for the parametric and non-parametric
methods. In Section \ref{robust}, we describe the results of the
simulations we have done to test the reliability of the
pipeline. Finally, we describe a multiprocessor implementation of
PyMorph and its performance characteristics (Section
\ref{parallel-sec}).

\section{Non-parametric methods in PyMorph}
\label{nonparam}

We have implemented an automatic procedure for structural
decomposition of galaxies using \gal supplemented by measurements
using non-parametric methods. The algorithms we use to estimate
non-parametric quantities are described in \nocite{con03}{Conselice} (2003) and
\nocite{lot04}{Lotz} {et~al.} (2004).  For completeness, we summarise the main features of
these algorithms. For the calculation of all non-parametric quantities
we use the sky value and center as determined by \sex, whereever
required. Also, we use the \sex source catalog to identify and mask
neighbour objects.

\subsection{Concentration index ($C$)}

Concentration is defined as the ratio of the radius of the galaxy which contains
80\% of the total light ($r_{80}$) to the radius of the galaxy which contains
20\% of the total light ($r_{20}$). ie.,

\begin{equation}
 C = 5 \log(\frac{r_{80}}{r_{20}})
\end{equation}
Here, the total light of the galaxy is taken to be the light within 1.5 times of
the Petrosian radius $r_p$ (hereafter extraction radius, $R_T$) where $r_p$ is
the radius of the galaxy at which the Petrosian parameter $\eta$ takes a value
of 0.2. The Petrosian parameter is defined as follows:

\begin{equation}
 \eta = \frac{\langle I_r \rangle}{\langle I \rangle_r}
\end{equation}
where $\langle I_r \rangle$ is the average light at the radius $r$ and $\langle
I \rangle_r$ is the average light inside $r$. Near the center of the galaxy,
where the light profile changes rapidly, we need to oversample the pixels to
obtain an accurate measurement. We achieve this by subpixelisation inside the
central 7 pixels (radius) of the galaxy by a factor of ten.

\subsection{Asymmetry ($A$)}
\begin{figure}
 \centering
 \includegraphics[scale=0.6]{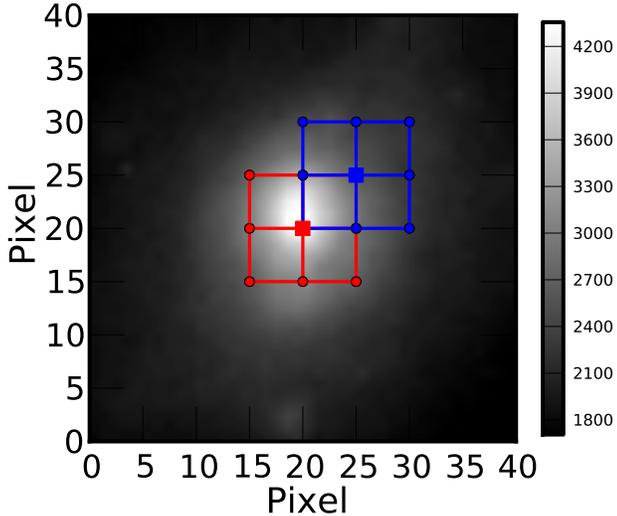}
 \caption{The centering correction in asymmetry measurement applied to galaxy
NGC 5585. The red square shows the initial estimate of the centroid of the
galaxy and the red circles represent the grid points. The blue square
represents the new centroid and the blue circles  correspond to the points on
the new grid. The grids are magnified 5 times, for illustration.} 
\label{asym-center}
\end{figure}

We calculate asymmetry using the algorithm devised by \nocite{con03}{Conselice} (2003). The steps
are: we rotate the galaxy through 180 degrees about its center which is taken to be the centroid (first image moment) of the galaxy pixels. We use bilinear interpolation to obtain the rotated
image. In the next step, we substract the rotated image from the original image
of the galaxy. From the residual image we estimate the total residual flux
inside the extraction radius. This is then normalised by the total flux of the
galaxy. This step can be represented as follows:
\begin{equation}
 A_O = \frac{\sum\left| I_0 - I_R\right|}{\sum I_0}
\end{equation}
where $I_0$ and $I_R$ are original image and the rotated image respectively and  the summation is over all valid pixels excluding the
pixels contaminated by light from neighbouring objects, inside the radius
$R_T$. In the next step, we identify possible biases in the measured
asymmetry value and correct for them. The first bias to the estimated
asymmetry value is due to the uncertainty in the estimated centroid of
the galaxy. For example, a perfectly symmetric galaxy can give
non-negligible asymmetry value if we cannot determine the centroid of
the galaxy exactly.  To correct for this bias we minimise the
asymmetry with respect to the center. To do that, we create a square
grid of nine points which includes the initial centroid of the
galaxy. The bin width of the grid of points is a fixed fraction of the
half light radius of the galaxy ($0.01 r_{50}$). 
We find asymmetry of the galaxy about all these
nine points, and assign the point corresponding to the minimum
asymmetry as the new centroid of the galaxy. We then generate a new
grid of points described as above, find the asymmetry of the galaxy
about this newly created grid of points and again determine the
minimum asymmetry. This process continues until we reach a stable
point where the asymmetry is minimum about that point compared to the
estimated asymmetry of the galaxy about the neighbouring points. This
process is illustrated in Figure \ref{asym-center} \nocite{con03}({Conselice} 2003).

\begin{figure*}
 \centering
 \includegraphics[scale=0.4]{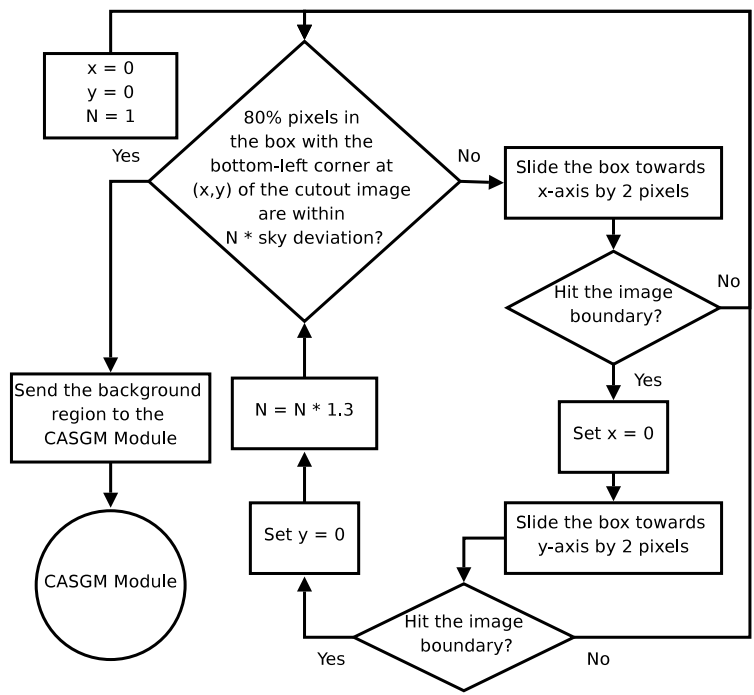}
 \caption{The algorithm to find empty background region within the galaxy
cutout. }
 \label{back-reg}
\end{figure*}

The second bias is introduced by the gradient of the sky near the
object.  To compensate for this, we estimate asymmetry of the sky and
substract it from the object asymmetry. To get the background
asymmetry we find an empty region near the object, which should
represent the real sky at the object position and therefore, cannot be
far from the object. On the other hand, it is difficult to get an
empty sky region very near to the object. This forces us to use an
optimum distance from the object center to search for the sky region
(This distance is same as the dimension of cutout of the object. We
explain this in section \ref{cutsize}). We mask all the objects in the
cutout and estimate the sky standard deviation ($\sigma_{sky}$).
Then, to identify a region of blank sky within the cutout, we put a
square region at the bottom left corner of the cutout with a size
X. We find that $20\leq X \leq 30$ pixels is optimal to get background
asymmetry. Inside the square region identified as possible blank sky,
we check how many pixels are within $\pm N \sigma_{sky}$. Initially,
we take N = 1. If we cannot find at least 80\% of the pixels within
this range we slide the square by 2 pixels along the x axis and repeat
the procedure. If we do not find a suitable blank sky region, this
process continues until we hit the image boundary. Then we slide the
blank sky search square along the y axis. If we fail to find a region
after searching the whole image, we assume that the sky has a large
gradient.  Therefore we increase N by 1.3 times and search sky region
inside $\pm N \sigma_{sky}$ as we did earlier. This simple approach
may fail when we have a highly crowded field. In such situations,
PyMorph notifies the user of the problem by setting the proper flags. This process is summarised in Figure \ref{back-reg}.

After we find an empty sky region we mask all pixels with counts
outside $\pm N \sigma_{sky}$ and find the asymmetry of that region and
minimize it in the same way as we did in the case of the
object. Finally, we subtract the minimised background asymmetry from
the minimised object asymmetry to get the 'true' asymmetry which can
be represented as:

\begin{equation}
 A = \min(A_O) - \min(A_B)
\end{equation}
where $A_O$ is the composite asymmetry of both object and sky, $A_B$ is the
asymmetry of the sky and $A$ is the true asymmetry \nocite{con03}({Conselice} 2003).

\subsection{Clumpiness ($S$)}
The clumpiness $S$ is a quantitative measure of clumpy regions in the
galaxy.  These are associated with star forming regions and thus
clumpiness of spiral galaxies is larger, on average, than that of
elliptical galaxies. To find $S$ we convolve the galaxy image with a
boxcar function of width $0.25 r_p$ where $r_p$ is the Petrosian
radius of the galaxy. This smoothed image is then subtracted from the
original image and the residual is summed within the extraction
radius. During this process, we mask the central part of the galaxy as
those regions are unresolved. The output of this process is the sum of
the clumpiness of the object and background. To get the clumpiness of
the object alone, we find the background clumpiness and substract it
from the composite value. The background region used for this purpose
is same as that used for the asymmetry calculation. The whole process
can be summarised in the following equation:

\begin{equation}
 S = 10 \left[\frac{\sum I_0 - I_S}{\sum I_0} - S_B\right] 
\end{equation}
where $I_0$ is the original image, $I_S$ is the smoothed image and the
summation is over all the positive pixels of residual image with the
annular region of width $0.2 R_T \leq r \leq R_T$. 
$S_B$ is the clumpiness of the background region \nocite{con03}({Conselice} 2003).

\subsection{Gini coefficient ($G$)}

It has been found that the Gini coefficient $G$ is a powerful way to
describe the morphology of a galaxy \nocite{lot04}({Lotz} {et~al.} 2004). This coefficient
can be regarded as a generalized concentration parameter. If all the
light in the galaxy belongs to a single pixel, the Gini coefficient
takes a value of 1. On the other hand, if the total light distributes
uniformly among all the pixels belongs to the galaxy, then the Gini
coefficient will be 0. On average, an elliptical galaxy has a
larger Gini coefficient than a disk galaxy.

To get $G$ we need to find the pixels in the image which belong to the
galaxy, i.e. obtain the segmentation map of the galaxy. This is
important as $G$ will be underestimated if we include sky pixels and
will be overestimated if we miss the outer pixels of the galaxy. To
determine which pixels belong to a galaxy, we convolve the galaxy
image with a boxcar filter of size $r_p/5$. This process will increase
the signal to noise ratio in the outer parts of the galaxy. Then we
measure the surface brightness $I_p$ at $r_p$. We assign all the
pixels in image with $I_p \leq I \leq 10\sigma$ to the galaxy where
$\sigma$ is the standard deviation of the sky. The upper limit ensures
that no cosmic rays or spurious pixels are included in the
segmentation map. Then the pixels belonging to the segmentation map
are sorted according to their photon count $I_i$ and $G$ is calculated
using the equation:

\begin{equation}
 G = \frac{1}{\bar{I_i} n (n - 1)} \sum_{i = 1}^{n} (2 i - n -1) |I_i|
\end{equation}
where $I_i$ is the photon count in the pixel  $i$ which belongs to the
segmentation map, $\bar{I_i}$ is the mean of all the pixel values $I_i$ and $n$
is the total number of pixels \nocite{lot04}({Lotz} {et~al.} 2004).

\subsection{Second order moment of the brightest pixels ($M20$)}

This quantity gives an idea of how the brightest pixels are
distributed over the galaxy segmentation map. For elliptical galaxies,
the brightest pixels are concentrated near the center of the
galaxy. Therefore, the $M20$ parameter will be smaller for ellipticals
compared to spiral galaxies where we observe large number of star
forming regions distributed all over the galaxy.  To compute $M20$ we
use the segmentation map generated to estimate the Gini
coefficient. We start by computing the flux weighted second order
moment of the galaxy $M_T$ as

\begin{equation}
 M_T = \sum M_i = \sum I_i \left[ (x_i - x_c)^2 + (y_i - y_c)^2\right] 
\end{equation}
where $I_i$, $x_i$, $y_i$ are the flux value and x and y coordinates
of the $i^{th}$ pixel in the segmentation map and $x_c$ and $y_c$ are
the initial center of the galaxy. We then minimize $M_T$ with respect
to the center of the galaxy as the initial value of the center is the
centroid of the galaxy. In the next step we sort the pixels according
to their flux value and find the moment of the 20\% brightest pixels
of the galaxy using the equation

\begin{equation}
 M20 = \log \left( \frac{\sum M_i}{M_T}\right) 
\end{equation}
where the summation continues until it satisfies $\sum I_i \leq 0.2
I_T$ where $i$ is the pixel in the sorted array of the segmentation
map and $I_T$ is the total light of galaxy \nocite{lot04}({Lotz} {et~al.} 2004). It can be seen
that when calculating $M20$ the pixels are weighted by $r^2$ which results in a
large $M20$ value for galaxies with many star forming regions distributed away
from its center. Therefore, $M20$ will be smaller for passbands which map the
underlying old stellar distribution (e.g. near IR) than those which map the young stellar population (e.g. near UV).

\section{Estimating structural parameters of galaxies}
\label{param}

Many well tested codes are available to perform 2D
decomposition. These include FITGAL \nocite{wad99}({Wadadekar} {et~al.} 1999), GIM2D
\nocite{sim98}({Simard} 1998), \gal \nocite{pen02}({Peng} {et~al.} 2002), BUDDA \nocite{sou04}({de Souza} {et~al.} 2004) etc. Although
the basic working principles of these codes are the same, they
implement different minimisation algorithms. FITGAL uses the
Davidon-Powell-Fletcher minimisation algorithms as implemented in the
{\it minuit} code developed at CERN \nocite{jam94}({James} 1994) while GIM2D
minimisation uses the Metropolis Algorithm. Marquardt-Levenberg
minimization drives \gal and BUDDA uses a multidimensional downhill
simplex method \nocite{pre92}({Press} {et~al.} 1992). Because of the complexity of the
parameter space it is very important to carefully setup the
minimisation so that these codes find the global minimum, as far as
possible. In PyMorph, we have chosen to use \gal for 2D decomposition
because of its simplicity and faster convergence.  However, the
pipeline is designed in a modular way so that the minimisation engine
can be easily changed at a future date, if required.

\begin{figure*}
 \centering
 \includegraphics[scale=0.7]{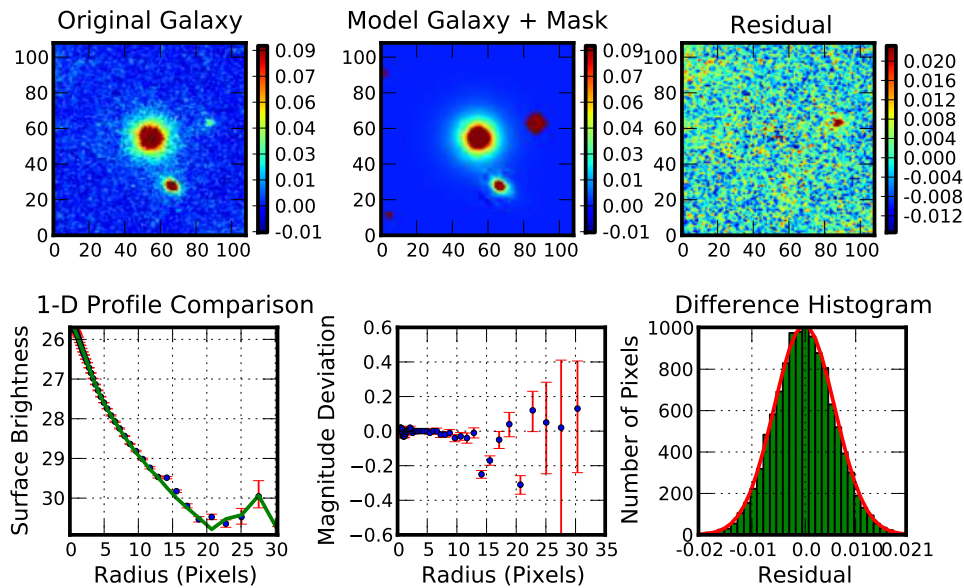}

\caption{Diagnostic output from PyMorph. The top left panel shows the image
  of the galaxy, top middle panel shows the best fit model image and
  the top right panel shows the residual (difference between galaxy and
  model) image after the fit. Lower left panel shows the one
  dimensional profile comparison of original (as data points) and
  model (as a solid line) for the galaxy. The lower middle panel shows
  the difference between the 1-D profiles of input and model galaxy. The
  lower right panel shows the histogram of the residual image, with the
  best fit Gaussian overplotted in red.}

\label{example}
\end{figure*}

The main preparatory steps before running \gal are: 1. Detect objects
in the input image and obtain their photometric parameters using \sex;
(again, the pipeline can be easily modified to use another source
extraction software) 2.  Create a cutout of the main object; 3. Create
an appropriate mask image to reject neighbour objects and spurious
pixels; 4. Create a configuration file for GALFIT. Besides 2D fitting,
PyMorph also performs the following tasks for every galaxy that it
fits: 1. Generates a one dimensional profile of input and best fit
model galaxy using the IRAF/STSDAS {\it ellipse} task to facilitate visual checks for obvious
fitting errors; 2. Converts all the fitted parameters to physical
parameters using the user specified cosmology and the galaxy redshift (if
available) 3. Creates outputs in several formats which includes CSV and
html and stores results in a \textit{mysql} database; 4. Creates
diagnostic plots in png format (see example in Figure \ref{example}).

\subsection{Object detection and photometry}

This is the initial stage of PyMorph. We use \sex to detect objects in
the input image and perform photometry on them. The input image may
either be a large frame or a cutout of the galaxy of interest. If the
image is a large frame, the astronomer may be interested only in a few
specific objects in the frame or may want to generate parameters for
all galaxies in the frame that satisfy some selection criteria. In
either case, before proceeding, a catalog with the exact location and
magnitude of the objects of interest is needed. To obtain such a
catalogue, PyMorph runs \sex on the input image to generate a \sex
photometric catalogue. The \sex output parameters used by PyMorph are
X\_IMAGE, Y\_IMAGE, ALPHA\_SKY, DELTA\_SKY, FLUX\_RADIUS,
THETA\_IMAGE, A\_IMAGE, ELONGATION, ISO0, BACKGROUND, CLASS\_STAR and
MAG\_AUTO. X\_IMAGE and Y\_IMAGE are the x and y coordinates
  respectively of the centroid of the object in pixel
  units. ALPHA\_SKY, DELTA\_SKY are the RA and DEC of the object.
FLUX\_RADIUS is the half light radius. THETA\_IMAGE, A\_IMAGE,
ELONGATION represent the position angle, the semi-major axis length
(a) and ratio of the semi-major to semi-minor axis length of the
object. \sex divides the detected objects into eight isophotal levels
above the ANALYSIS\_THRESH. ISO0 represents the area of the object
above the ANALYSIS\_THRESH in units of pixel$^2$.  MAG\_AUTO,
BACKGROUND and CLASS\_STAR are the magnitude, background value at the
object position and the stellarity parameter of the object. The user
can choose the \sex local or global background. CLASS\_STAR$=\sim1$
for a star and $\sim0$ for a galaxy.

\subsection{Position match and object cutout generation}
\label{cutsize}

After the \sex catalogue is created, the program compares the user
given input catalogue of objects (which lists the galaxies of
interest) with the \sex catalogue. For each object in the input
catalogue, PyMorph finds the corresponding entry in the \sex catalogue
either by matching the RA and DEC coordinates or by matching pixel
coordinates. The matching radius can be set by the user, either in
units of pixels or in arcsec. For every successful match between the
input catalog and the \sex catalogue, PyMorph reads all the required
\sex photometric parameters of that object for further use.

The next step is to create a cutout image of the object to feed to
GALFIT.  The size of the cutout image should be such that it contains
enough sky pixels without becoming very large in size. The first
criterion is highly desirable as insufficient number of sky pixels may
cause incorrect background estimation by GALFIT. This will seriously
affect the estimation of S\'ersic index of the bulge.  On the other
hand, including a large sky region in the cutout will increase
computational resource requirements. We use the \sex FLUX\_RADIUS
($R_{50}$), ELONGATION ($a/b$) and THETA\_IMAGE ($\theta$) parameters
to find the optimum size of the cutout as follows:

\begin{eqnarray}
X &=& F_{\textrm{rad}} R_{50} \left(|\cos\theta| + \frac{b}{a} | \sin\theta
|\right) \nonumber \\
Y &=& F_{\textrm{rad}} R_{50} \left(| \sin\theta | + \frac{b}{a} | \cos\theta
|\right)
\label{size-eqn}
\end{eqnarray}

where $X$ and $Y$ are the dimension of the cutout centered on the galaxy
center and $F_{\textrm{rad}}$ is a user specified parameter. Through
trial and error, we found that $F_{\textrm{rad}} = 6$ gives optimum
size for the cutout image. Using the size and the centroid parameters
of the object, we cut the portion of image and the corresponding
weight image (if available).

\subsection{Create mask image}

\begin{figure*}
 \centering
 \includegraphics[scale=0.4]{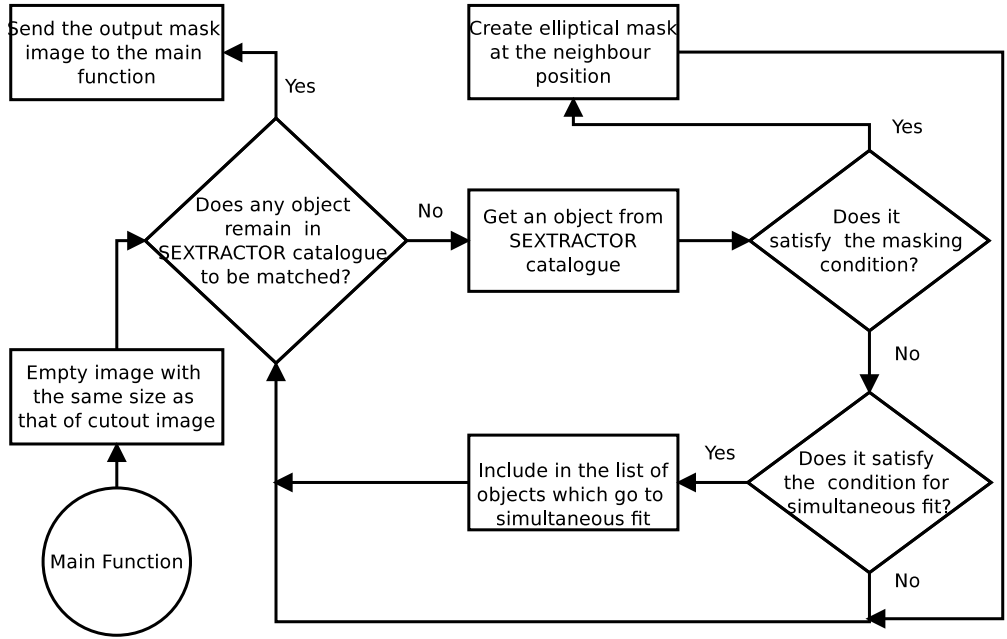}
 \caption{The algorithm to create the mask image.}
 \label{mask}
\end{figure*}

An advantage of \gal is that it allows us to simultaneously fit any
number of objects. But it is not advisable to fit many objects
simultaneously as that increases the number of free parameters.  In
such situations, the fit may converge to a local minimum. Therefore,
we should simultaneously fit for a neighbouring object only if it is
large and/or bright enough to significantly contaminate the main
object. To decide whether a neighbouring object should be included in
the simultaneous fitting, we use the A\_IMAGE, ISO0 parameters of each
object. We compare the parameters of all neighbouring objects in the
\sex catalogue with those of the object of interest. We check whether
the objects overlap with the main object by comparing their scaled
semi-major axis (the scale is user specified). The scale
  determines the distance to the closest neighbour to be fitted
  simultaneously with the object. For e.g., let us say that the
  distance between the centers of the object of interest and its
  neighbour is 100 pixels, the semi-major axis of the object of is 60
  pixels and that of the neighbour is 30 pixels. Then, even if the
  galaxies have circular shape they will not overlap (because $ 60 +
  30 < 100$) and the neighbour will therefore be masked. If the user
  requires that such neighbours be fitted simultaneously, the scale
  parameter can be tweaked. If the scale parameter is set to 2, then
  $(60 + 30) \times 2 > 100$ and now the neighbour will be fitted
  simultaneously with the object. If the program finds overlapping
neighbours, it also checks whether the area of the neighbour is larger than
a user specified fraction of the main object. If the area (ISO0) of
the neighbour is higher than a threshold fraction of the area of the
object of interest, then it will be fitted simultaneously. If the
neighbour's area is below the threshold fraction, then that neighbour
will be masked irrespective of its distance from the object. Therefore
the condition for simultaneous fitting is as follows:

\begin{eqnarray}
d  &<& T_R (R_o + R_n) \quad \textrm{and} \nonumber \\
A_n &>& T_A A_o \textrm{,}
\label{mask-eqn}
\end{eqnarray}
where $d$ is the distance between object and neighbour, and $R_o$,
$R_n$ are the A\_IMAGE parameters and $A_o$ and $A_n$ are the ISO0
parameters of object and neighbour respectively. $T_R$ is a user
specified parameter which decides the threshold fraction of overlap
between object and neighbour. The value of $T_A$ decides the smallest
object which is to be included in the simultaneous fit with the object
of interest. We found the control parameters $T_R = 3.0$ and $T_A =
0.3$ give good results. Neighbours which do not satisfy the
simultaneous fit criterion in Eqn \ref{mask-eqn} are masked out. To do
this, PyMorph creates an elliptical mask at the position of the
object.  This mask has the same ellipticity as that of the neighbour
but its semi-major axis is scaled by an amount $T_M$ which is also to
be specified by the user. i.e., the semi-major axis of the mask
becomes $T_M R_n$. This process continues for all the objects in the
\sex catalogue and finally we have a mask image that is the union of
masks for all the contaminating objects. The block diagram describing
this process is shown in Figure \ref{mask}.

This masking technique works only if SExtractor detects the
neighbouring object, in the first place.  Since that process depends
on the DETECT\_THRESH and DETECT\_MINAREA parameters the user chooses,
it is possible that some spurious pixels bright enough to affect the
fit may be left undetected by SExtractor.  So, we use the following
simple technique to mask such pixels.  From the center of the main
object, we make elliptical annuli with increasing radii.  In the inner
aperture we find the maximum value of the galaxy.  We assume a smooth
light distribution for the galaxy which decreases with distance from
the center.  This implies that the largest value in the central
elliptical aperture is the maximum value the object can have.  So, we
mask all the other pixels outside the inner aperture with value
greater than this maximum value.  Now we go to the next pair of annuli
and find the maximum and mask other pixels outside this aperture which
have value larger than the maximum of this aperture.  This procedure
continues till the aperture radius hits the image limit. 
  Using this technique we are likely to mask small regions with a high
  flux e.g. knots of star-forming regions in spiral arms. However,
  since we are attempting to determine global parameters for the bulge
  and disk of the galaxies, masking out local fluctuations over the
  underlying bulge and disk will likely only improve the parameter
  estimation. However, if desired, one can switch
  off this masking technique setting the 'mask-norm' option. In that
  case, only the neighbour objects will be masked. Now, almost all
the spurious pixels which may be part of undetected objects should be
masked correctly. We combine this mask image with the mask image made
using the \sex catalogue to get the final mask.

\subsection{Create configuration file for \gal}

\gal configuration can be done either through a text file or through the command
line. Pymorph creates an input configuration file for each object in the user given
catalogue to feed GALFIT. This file specifies the filenames of the input image,
the weight image and the point spread function image. The other entries include
 initial values of the parameters of the  components used in the fitting.
For two dimensional decomposition, a galaxy is usually assumed to have, at
least, a bulge and a disk component. The bulge  is modeled by a S\'ersic
function of the form

\begin{equation}
I(r) = I_e \exp\left({-b_n \left[\left(\frac{r}{r_e}\right)^{1/n} -
1\right]}\right) 
\end{equation}
where $I(r)$ is the intensity of the bulge at radius $r$,$I_e$ is the intensity of the bulge at radius $r_e$, 
$r_e$ is the half light radius and $n$ is the S\'ersic index of the
bulge. $b_n$ is a quantity which depends on $n$. Similarly, the disk
part is usually modeled using the exponential form

\begin{equation}
 I(r) = I_d \exp\left({-\frac{r}{r_d}}\right)
\end{equation}
where $I_d$ is the disk intensity at the centre and $r_d$ is the disk
scale length.  The surface brightness profile of the galaxy is modeled
as a linear combination of these two functions.

\begin{figure*}
 \centering
 \includegraphics[scale=0.4]{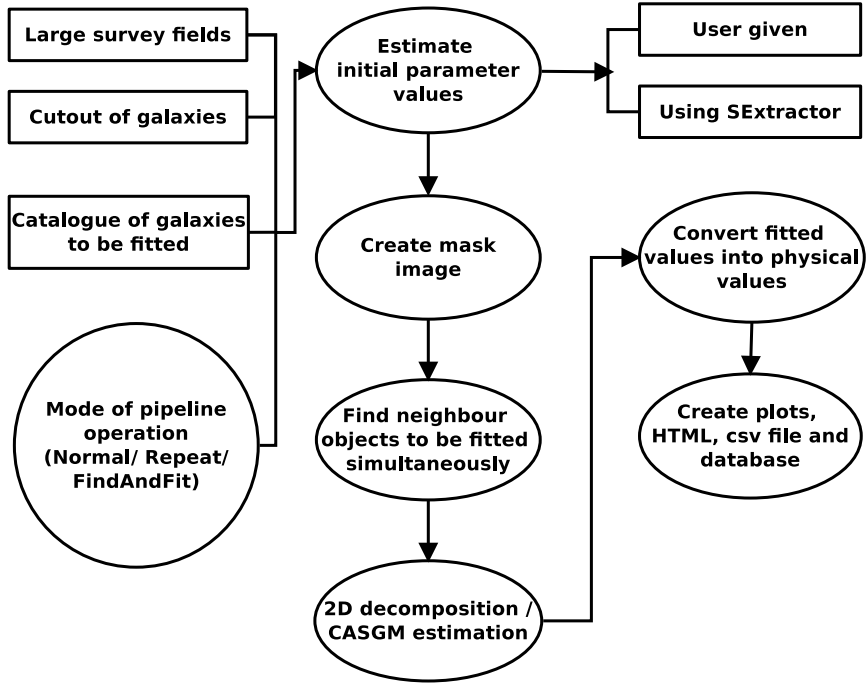}
 \caption{Flowchart of PyMorph}
 \label{flowchart}
\end{figure*}

There are more than a dozen parameters to be fitted during the
decomposition of a galaxy. These include the centers of bulge and disk
components and their total magnitudes, scale radii, axis ratio and
position angles.  S\'ersic index of the bulge is another parameter
involved in the fitting. \gal offers two additional parameters which
model the boxiness/diskiness of the bulge and disk. We have not
used these parameters in our fits.
  Apart from the parameters involved in the photometric components of the
galaxy, one other important parameter is the sky. There is an option in \gal to
fit sky with gradient in x and y directions of the image.  To increase our
chances of finding the global minimum from \gal, we need to set the appropriate
initial values for the fit parameters. We use \sex MAG\_AUTO and FLUX\_RADIUS
parameters of the galaxy as the initial values for the total magnitudes and
scale radii of both bulge and disk. The initial values of the axis ratios of
both bulge and disk are set from the \sex ELONGATION parameter and position
angle is calculated from THETA\_IMAGE. We always set the initial value
for the S\'ersic index $n$ to 4.  The sky parameter is set to the \sex
value. We found that \sex slightly overestimates the background value
which can result in incorrect estimation of bulge parameters. This issue will be
discussed in Section \ref{caveats}. The working of PyMorph is summarised as a
block diagram in Figure \ref{flowchart}.

\section{Setting up PyMorph}
\label{setup}

PyMorph is written entirely in the Python programming language. Python
is a modern, high level programming language with many features that
encourage readable and reusable code. Several current astronomical
data analysis systems have been made accesible through Python
(e.g. Pyraf for IRAF and CASA.py for AIPS++). Python will play a major
role in many new software initiatives in astronomy.

Besides the standard python modules, additional modules required by
PyMorph are numpy, matplotlib and pyfits. Numpy is used for arithmetic
manipulations on the image array and matplotlib is used to generate
output plots. Pyfits is a python module to read and write fits
images. As we have already mentioned, \sex and \gal currently serve
PyMorph as the detection and fitting programs. Since all the required
processing for 2D decomposition is pipelined, the user needs to
initialize the code correctly so that it completes without needing
further intervention. All the required input parameters can be set
through an input configuration file. Some parameters can be give to
PyMorph via command line as well. An example input configuration file
for PyMorph is shown in Figure \ref{config}.

\begin{figure*}
\begin{center}
\begin{verbatim}
(A) ###----Specify the input images and Catalogues----###
imagefile = 'j8f643-1-1_drz_sci.fits'
whtfile = 'j8f643-1-1_drz_rms.fits'   #The weight image. 
sex_cata = 'j8f643_sex.cat'           #The sextractor catalogue 
clus_cata = 'cl1216-1201.cat.old'     #The input catalogue of galaxies
(B) ###----Specify the output images and catalogues----###
out_cata = 'cl1216-1201_out.cat'      #catalogue of galaxies in the field
rootname = 'j8f643'
(C) ###----Point spread function list----###
psflist = '@psflist.list'             #List of psf stars
mag_zero = 25.256                     #magnitude zero point
(D) ###----Conditions for Masking----###
mask_reg = 2.0
thresh_area = 0.2
threshold = 3.0
(E) ###---Size of the cut out and search conditions---###
###---size = [resize?, varsize?, fracrad, square?, fixsize]---###
size = [0, 1, 6, 1, 120]              #size of the stamp image
searchrad = '0.3arc'                  #The search radius  
(F) ###----Parameters for calculating the physical parameters of galaxy----###
pixelscale = 0.045                    #Pixel scale (arcsec/pixel)
H0 = 71                               #Hubble parameter
WM = 0.27                             #Omega matter
WV = 0.73                             #Omega Lambda
(G) ###----Parameters to be set for calculating the CASGM----###
back_extraction_radius = 15.0
angle = 180.0
(H) ###----Fitting modes----###
repeat = False                        #Repeat the pipeline manually
galcut = False                        #True if user provides cutouts
decompose = True                      #Find structural parameters
cas = True                            #Find CASGM parameters
findandfit = 0                        #Run for all objects which satisfies user
defined criteria
crashhandler = 0
(I) ###---Galfit Controls---###
components = ['bulge', 'disk']        #The components to be fitted to the object
###---fixing = [bulge_center, disk_center, sky]
fitting = [1, 1, 0]                    # = 0, Fix params at SExtractor value
(J) ###----Set the SExtractor and GALFIT path here----###
GALFIT_PATH = '/home/vinu/software/galfit/modified/galfit' 
SEX_PATH = '/home/vinu/software/sextractor-2.5.0/sex/bin/sex'
PYMORPH_PATH = '/home/vinu/ncra/vinucodes/serial_pipeline/trunk/pymorph'
(K) ###----The following conditions are used to classify fit as good/bad----###
chi2sq = 1.9                          #< chi2sq
center_deviation = 3.0                #< abs(center - fitted center)
(L) ###----Database Informations----###
database = 'cluster'
table = 'clusterfitresults'
usr = 'vinu'
pword = 'cluster'
dbparams = ['Cluster:cl1216-1201', 'ObsID:1:int']
\end{verbatim}
\end{center}
\caption{Sample input configuration file for PyMorph}
\label{config}
\end{figure*}

There are 12 blocks in the configuration file. Most of the parameters
in blocks A and B are self-explanatory. The parameter \textit{psflist}
corresponds to a file containing the list of suitable PSF images.  These may be stellar images from the input frame (typically
  unsaturated, isolated bright stars) which the user feels are
  accurate representations of the PSF. It is known that in large
frames the PSF may vary spatially. Therefore the general principle is
to use the nearest star to the object as the PSF. If the names of
these stellar images follow the convention \textit{psf\_RaDec.fits},
eg. \textit{psf\_1216382-1200443.fits}, then PyMorph will
automatically find the nearest PSF to the object from \textit{psflist}
and use it for fitting, otherwise it assumes that there is a one to
one correspondence between the list of PSFs and galaxies to be fitted.
The \textit{thresh\_area} and \textit{threshold} parameters in block D
correspond to $T_R$ and $T_A$ respectively in
Eqn. \ref{mask-eqn}. \textit{mask\_reg} ($T_M$) determines how much
area of the neighbour should be masked.

The \textit{size} keyword is a list of five parameters which control
the size of the cutout image. The first one in the list is a flag that
tells the program to create a cutout of the object. This is needed
sometimes when the user already has a cutout of the objects. In those
situations, this parameter should be unset. Using the second entry in
the list, the user can set the size of the cutout of all the objects
fixed at a particular value irrespective of the real size of the
object. If the user wants to create cutout of objects scaled by their
angular size, the size of the cutout will be determined by the third
parameter in the list. This is the quantity $F_{rad}$ involved in 
Eqn. \ref{size-eqn}. The fourth quantity determines shape of the
cutout. If the user sets the parameter, then the cutout will have a
square shape with size equal to the maximum of $X$ and $Y$ which is
given by the Eqn.  \ref{size-eqn}. Otherwise the cutout will have
rectangular shape of size $X$ and $Y$. The final entry in the list is
used if the user does not set the second entry. e.g. an entry
\textit{size = [1, 1, 6, 1, 120]} tells the program to create a
cutout, measure size from the objects, that the size should be six
times larger than the half light radius of the objects and to make a cutout
of square shape. On the other hand, if the user sets the size keyword as
[1, 0, 6, 1, 120], then the program will create a cutout of size $120 \times
120$ irrespective of the size of the galaxy.

The parameter \textit{searchrad} is the search radius used to match the input
catalogue with the \sex catalogue. The comparison can either be in
pixels or in arcsec. Therefore if the user set \textit{searchrad =
  '0.3arc'} then the program matches objects within 0.3 arcsec. On the
other hand, the entry \textit{searchrad = '5pix'} matches objects
within 5 pixels. The cosmological parameters in the block F will be
used to convert the fitted parameters to physical units. The G block
parameters will be used for the calculation of asymmetry of
objects. Through \textit{back\_extraction\_radius} parameter the user
can set the size of the background region which will be used to find
the asymmetry of the background.  The parameters in block H determine
the different modes of PyMorph behaviour.  \textit{galcut} should be
set if the user wants to supply cutouts of galaxies.
\textit{decompose} and \textit{cas} should be set to get parametric
and non-parametric results respectively, from PyMorph. The
\textit{findandfit} is implemented in order to get the structural
parameters of all the objects in large frame(s) which satisfies some
user defined criteria. For example, this mode can be used if the user
wants to generate the structural parameters for all objects in a
frame between magnitudes $m_1$ and $m_2$. If \textit{crashhandler} is
set then the program automatically reruns \gal for some obvious
fitting errors.
In the presence of a bright neighbour or dust
lanes in the galaxy, the best fit center of either bulge or disk can
be significantly different from the optical centroid of the
galaxy. Pymorph tracks errors at all the stages of the pipeline and
saves those errors as flags. Therefore it is possible to identify the
stages of the pipeline that have failed, in some way. The obvious
fitting errors include incorrect centers, ie. the fitted center of the
component is very far from the visible center of the galaxy, or some
parameters have hit the limit of their allowed range. After a fitting
process finishes, pymorph checks for these errors. If the program
finds an incorrect center, the galaxy is refit with tight constraints
on the range of center. If the program finds a parameter that hits the
limit, then it increases the fitting range and carries out a refit, which
may give a better result. If the refit also fails, the galaxy is flagged as a
poor fit (indicated in the FIT parameter in the output catalogue).
Finally, the \textit{repeat} mode is implemented to rerun
PyMorph semi-automatically. It is possible that \gal converges to a
local minimum for a few galaxies. Then the user can edit the
corresponding \gal configuration file, mask etc. to rerun \gal again,
if the pipeline is set to run in \textit{repeat} mode. Pymorph generates all
the necessary intermediate files to do decomposition which includes
the mask, \gal configuration files etc. and saves them to disk. If the
user finds that \gal failed to converge to a global minimum because of
improper initial values of the parameters or a poor mask image, then the user
can manually edit the mask image or slightly alter the \gal initial
values. After this, the user can set the repeat mode and run
pymorph. In that case, pymorph uses the existing intermediate files to
run \gal without generating them anew.

The I block parameters controls GALFIT. Through \textit{components}
keyword the user can set the number of photometric components of
galaxies for fitting. If \textit{components = ['bulge', 'disk']}, then
a S\'ersic and an exponential function will be fitted to the galaxy's
surface brightness. Through \textit{fitting} the user can fix/free the
centers of the fitting functions and sky. For example, \textit{fitting
  = [1, 1, 0]} tells the program to set the centers of the bulge and
disk as free parameters and fix sky at the initial value during
fitting. The program reads the location of \sex, \gal and PyMorph from
the J block. The K block parameters will be used to determine whether
a given fit is acceptable or not. This includes the simple reduced
$\chi^2$ (\textit{chisq}) and
\textit{center\_deviation}. \textit{center\_deviation} is a measure of
the difference between the initial and fitted centers of the
components in pixel units.
If this difference is greater than
\textit{center\_deviation}, then the corresponding fit will be
considered as bad. This will be refitted with tight constraints on the
centers provided the user has set \textit{crashhandler} parameters in
the H block. The final L block deals with the output database. If the
program finds a \textit{mysql} database server then these parameters
become active.  All other parameters in this block are self
explanatory other than \textit{dbparams}. Through \textit{dbparams}
the user can create additional columns in the database table and set
their values. For example, \textit{dbparams = ['Cluster:cl1216-1201',
    'ObsID:1:int']} will create two additional columns Cluster and
ObsID in the current database table. The functions of different blocks
in the PyMorph configuration file are summarised in Table
\ref{tab-blocks}

\begin{table}
 \centering
  \caption{Functions of different blocks in  the PyMorph configuration file.} 
  \label{tab-blocks}
  \begin{tabular}{@{}lc@{}}
  \hline
Block & Function \\
 \hline
A & Input images and catalogues\\
B & Output files\\
C & PSF file\\
D & Masking conditions\\
E & Cutout size\\
F & Cosmology\\
G & CASGM parameter measurement\\
H & Fitting Modes\\
I & \gal controls\\
J & Set path to software\\
K & Classification criteria\\
L & Database information\\
\hline
\end{tabular}
\end{table}

\section{Testing the robustness of PyMorph}
\label{robust}

\subsection{Caveats of pipeline use}
\label{caveats}
 
Before we describe the tests we carried out to examine the robustness of
the Pymorph pipeline we would like to caution the user against using
it blindly without accounting for its limitations. Specifically:

\begin{enumerate}
\item A linear combination of bulge and disk is inadequate to model
  galaxy structures such as a nuclear point source, bars, rings etc.,
  whenever they are sufficiently strong. Adding analytic models for
  these components greatly increases the free parameters in the
  minimisation making it more likely to converge to an incorrect local
  minimum. Automated procedures that attempt to fit all these
  components are unlikely to give reliable results. We have therefore
  deliberately not added the ability to fit additional components to
  Pymorph. \gal does provide for modelling these features, but one
  needs to run it carefully on individual galaxies. In
the rest frame near-infrared, components like star forming knots
are quite weak. In addition, for distant galaxies, these small scale features will be blended with the large scale bulge and disk. In such cases, a linear combination of bulge and disk will likely be a robust model of the galaxy structure, although its physical interpretation is more complicated.
\item Certain minimisation algorithms are more prone to converge to
  local minima (e.g. Levenberg-Marquardt used by \gal) than others
  (e.g. Metropolis algorithm used by GIM2D). In practice, \nocite{hau07}{H{\"a}ussler} {et~al.} (2007) have
shown that \gal works better than GIM2D in many situations. Users need to be 
aware of the capabilities and limitations of the algorithm and specific implementation
being used.
\item \nocite{hau07}{H{\"a}ussler} {et~al.} (2007) have pointed out that the \sex sky determination tends to
  overestimate the background. Incorrect background determination can
  affect parameter estimation, especially those of the bulge.
\item If one is fitting a pure disk galaxy, it often gets incorrectly
  fit by a S\'ersic function with $n=1$. This results in $B/T \sim 1$,
  which is clearly incorrect.
\end{enumerate} 

In order to assist the user in determining whether a particular galaxy
has been correctly fit, Pymorph provides a diagnostic plot (see sample
in Fig. \ref{example}). We recommend the following procedure to test
for the quality of the fit, which users may adapt to their
requirements.

\begin{enumerate}
\item Check whether the FIT parameter in the output catalogue is unset. If it is
unset it means that the reduced $\chi^2$ is larger than that the user specified
in the config.py or the fitted center of at least one component is incorrect.
\item Check for large residuals near the centre of the residual image
in the diagnostic plot, which may be caused by a wrong PSF.
\item Check whether the difference histogram is centered at zero and well 
matched to the best fit Gaussian. If not, the residual image is not purely
composed of noise. 
\item Check whether the one dimensional profiles of original galaxy and the
model galaxy match.
\end{enumerate}

Through experience, the user will be able to rapidly identify problematic fits. Some of the quality checks above can be automated via scripts that use the
information contained in the ASCII output files (\textit{result.csv}) produced
by Pymorph. Galaxies that are poorly fit may be handled using PyMorph in
repeat mode (see Section \ref{setup}).

\subsection{Compare extracted CASGM parameters with published values}

\begin{figure*}
 \centering
 \includegraphics[scale=0.4]{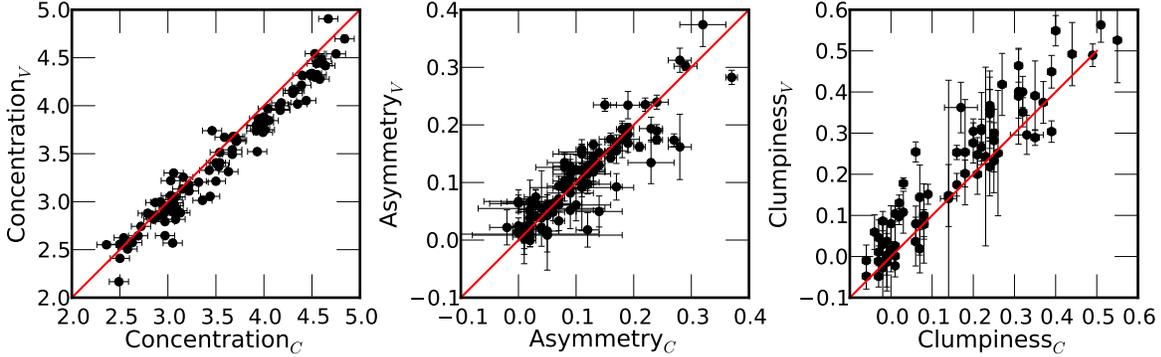}
 \caption{The comparison of CAS parameters from PyMorph with the value given by
\protect\nocite{con03}{Conselice} (2003). On the y-axis we show values estimated by PyMorph and on the
x-axis values from the published catalogue.}
 \label{vinu-conse}
\end{figure*}

\begin{figure*}
 \centering
 \includegraphics[scale=0.6]{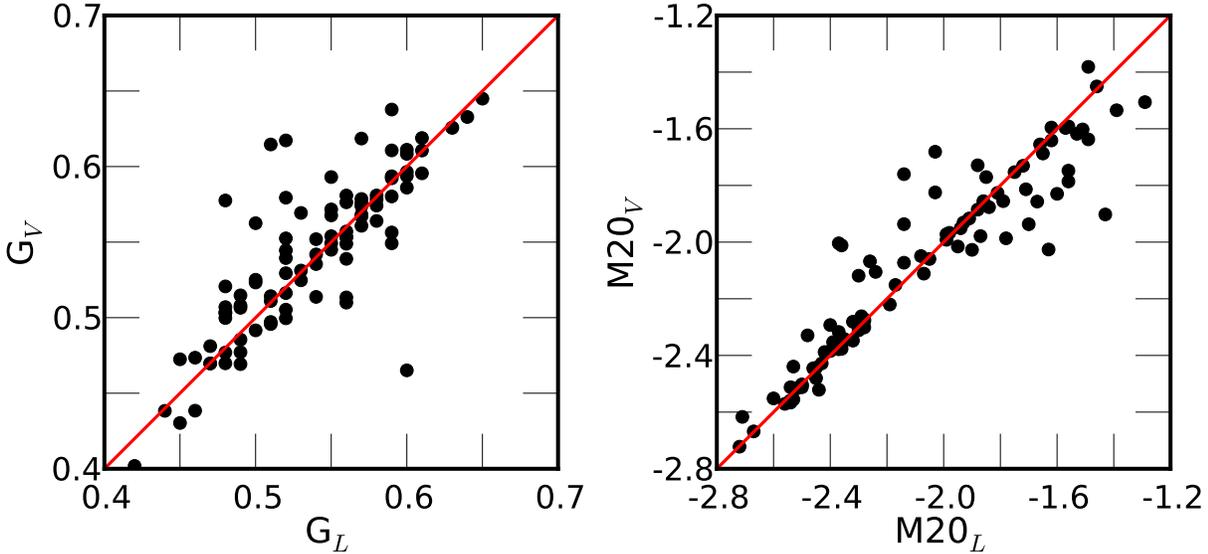}
 \caption{The comparison of GM parameters from PyMorph with the values given by
\protect\nocite{lot04}{Lotz} {et~al.} (2004). On the left panel we compare the Gini coefficients and on the
right panel we compare the second order moments. The y-axis shows the 
values measured by PyMorph and the x-axis values from \protect\nocite{lot04}{Lotz} {et~al.} (2004).}
 \label{vinu-lotz}
\end{figure*}

We have used a well studied sample of nearby galaxies \nocite{fre96}({Frei} {et~al.} 1996)
with publicly available data to test the robustness of the CASGM
parameters. We compare our estimated values with published values for
these galaxies by \nocite{con03}{Conselice} (2003) and \nocite{lot04}{Lotz} {et~al.} (2004). This gives an idea
of the robustness of our automated procedure. Figures \ref{vinu-conse}
and \ref{vinu-lotz} show the result of this comparison. We calculated
the dispersion between our values and the published values. We found
that the average deviation for concentration, asymmetry and clumpiness
from that of \nocite{con03}{Conselice} (2003) are -0.11 $\pm$ 0.14, 0.0 $\pm$ 0.036, 0.06
$\pm$ 0.09 respectively. The average dispersion of our estimated
values for Gini coefficient and second order moment with that of
\nocite{lot04}{Lotz} {et~al.} (2004) are 0.0 $\pm$ 0.035 and 0.0 $\pm$ 0.16. It will be
interesting to compare the CAS values of \nocite{con03}{Conselice} (2003) with those
\nocite{lot04}{Lotz} {et~al.} (2004) to show that similar dispersion was seen in previous comparisons. The dispersion between the CAS parameters reported in those papers are 0.08 $\pm$
0.16, -0.04 $\pm$ 0.0445 and 0.01 $\pm$ 0.08 for $C$, $A$ and $S$ respectively.

The small systematic offset of our concentration measurement
  from that of \nocite{con03}{Conselice} (2003) (see left panel of
  Figure~\ref{vinu-conse} is due to the slight error in measuring the
  background value.  This error will propogate to the measurement of
  $r_{20}$ and $r_{80}$ values.  This effect is stronger in the case
  of galaxies with high concentration as they will have smaller
  $r_{20}$ and the slight uncertainty in $r_{20}$ leads to some
  variation in concentration index. It should be noted that this
  offset, though real, is small and within the error in many cases.

\subsection{Simulating two dimensional galaxy light profiles}

To test the robustness of the structural parameters given by PyMorph
we have run the code on simulated galaxy light profiles. In this
section, we describe the simulation and then discuss the
results. We simulate surface brightness profiles of galaxies as a
linear combination of a S\'ersic and exponential functions. The steps
involved in the simulation are the following:

1. Set the values of the parameters involved in the S\'ersic and
exponential functions. The range of these parameters used for
simulation are 18 $<m_b <$ 25, 1 kpc $< r_e <$ 6 kpc, 1 kpc $< r_d <$
10 kpc, 0.4 $< e_b <$ 0.9, 0.2 $< e_d <$ 0.9 where $m_b$, $r_e$, $e_b$
are the apparent magnitude, scale radius and axis ratio of bulge
component of the galaxy and $r_d$, $e_d$ are the apparent magnitude,
scale radius and axis ratio of the disk component. The range of
S\'{e}rsic index used is $1 < n < 6.0$.  These parameters are
distributed uniformly along their respective ranges.

2. The S\'ersic function is steeper towards the center for large
values of S\'ersic index. Therefore, it is important to treat this
cusp differently to generate exact surface brightness in the central
region. We do this by oversampling the S\'ersic function at the center
region. We oversample the central $5\times5$ pixels by a factor of
10. At the central pixel, we oversample the function 30 times while
conserving the flux. For exponential function, the oversampling is
done with a factor of 10.

3. We simulated 1000 objects as described in the previous step
and inserted them into a $6000\times6000$ array. The position of these galaxy
profiles in the large array are distributed randomly. This process is
intended to simulate the original observation.

4. We then convolved the model image by the PSF.  We have extracted a
stellar image from an ACS observation for this purpose.

5. To simulate noise, we have generated a background image which has a
typical standard deviation of the original observations of HST
ACS/WFPC2. The background values are distributed according to the
Poisson distribution. Along with the background image we propagate the
statistical error of the object counts determined from model image
 to create the noise image.

6. The model image and the background images are added to get the
final simulated images and these images are used for further analysis.

\subsection{Regression results for simulated galaxies}
\label{regress}
\begin{figure*}
 \centering
 \includegraphics[scale=0.6]{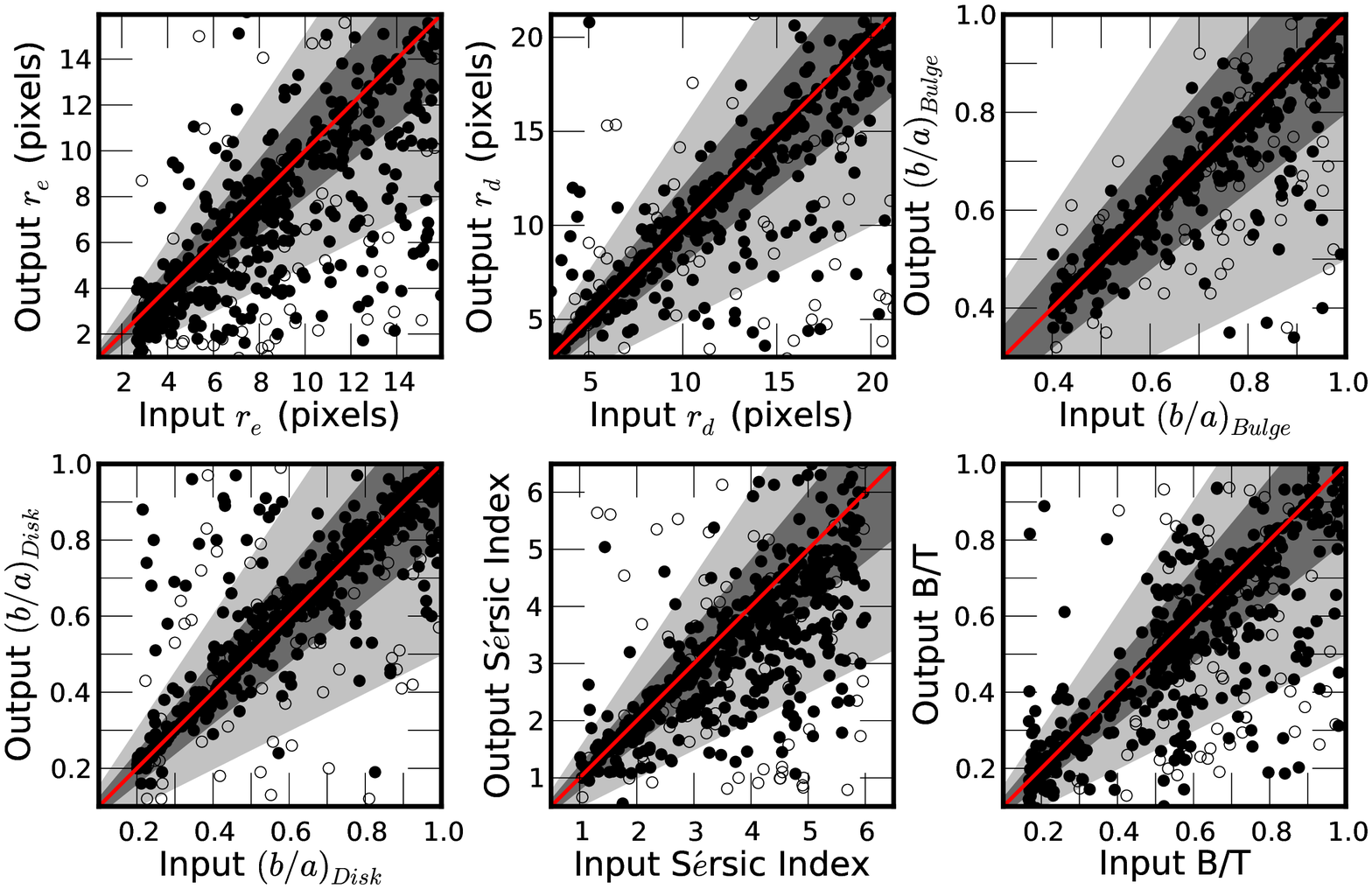}
 \caption{Input and recovered parameter values of galaxies at z =
   0.8 where we have converted the physical parameters of galaxies such
as scale lengths to pixel units using standard cosmology by assuming
that the galaxies are at z = 0.8. The dark grey represents the region
 in which the fractional
   error is 20\% and light grey represents the regions of 50\%
   fractional error. The filled circles are for galaxies with
   magnitude $< 23.0$ and open circles are for magnitude $> 23.0$.}
 \label{in-out8}
\end{figure*}

\begin{figure*}
 \centering
 \includegraphics[scale=0.6]{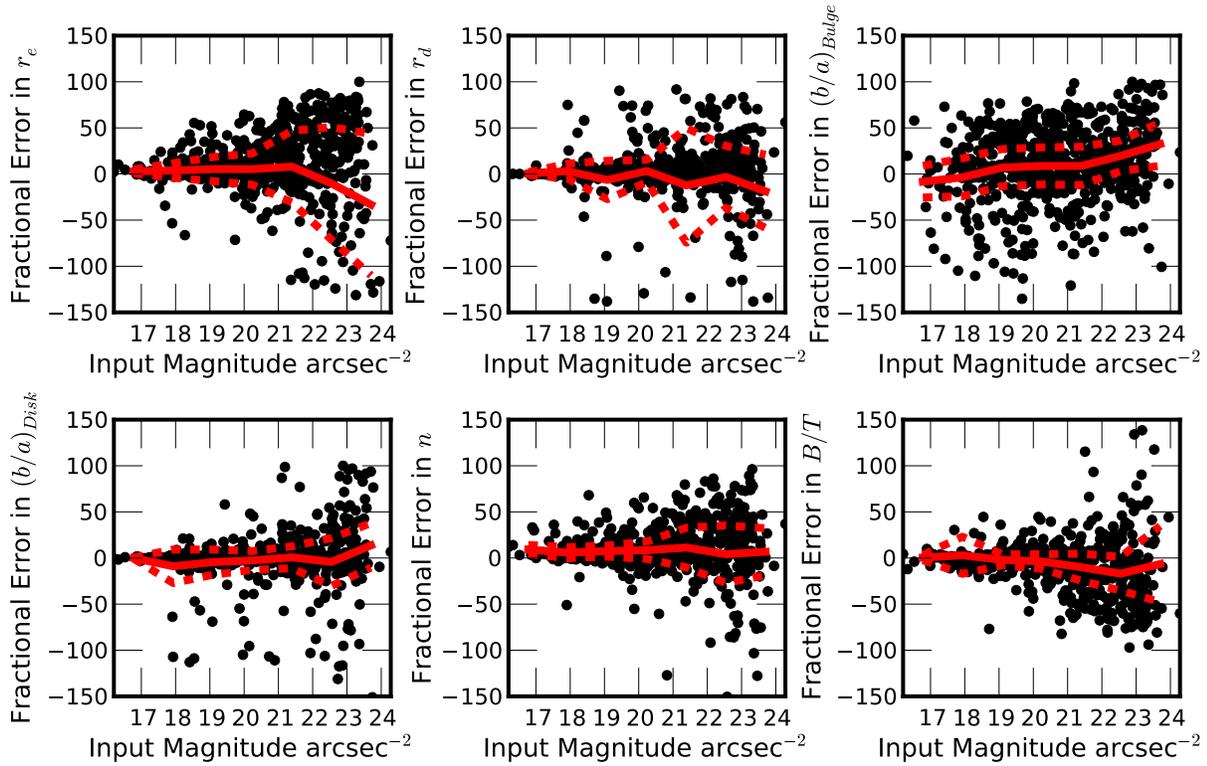}
 \caption{The fractional  error on the recovered parameter values of galaxies at
z = 0.8. The solid red line indicates the mean and the dashed red lines
represent the 1$\sigma$ region. The fractional error (in percent) is calculated as
(Output -Input) * 100 / Input.}
 \label{frac-err}
\end{figure*}

\begin{figure}
 \centering
 \includegraphics[scale=0.5]{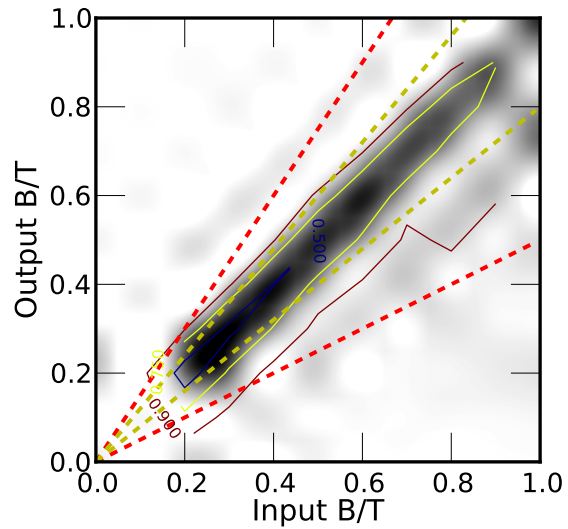}
 \caption{The contours which contain 50, 75 and 90\% of the
   galaxies. The region between thick dashed yellow lines represents
   20\% variation from the true value and the region within the thick
   dashed red line represents 50\% variation from the true value. Only
   the input $B/T$ vs output $B/T$ is shown here.}
 \label{frac-cont}
\end{figure}

We have used PyMorph to extract parameters from the simulated
images. The regression test results are shown in Figure
\ref{in-out8}. We have calculated the mean magnitude per arcsec$^{2}$
within the half light radius of the galaxy, which is given by the
SExtractor FLUX\_RADIUS parameter. We found that for most of the cases
the extracted parameters are within a fractional error of 50\% for a
mean surface brightness $< 23$ arcsec$^{2}$ (Figure  \ref{frac-err}). It is
found that for $\sim$ 75\% objects the recovered parameters are within 20\% of
the input value. Also for 90\% cases the recovered values are within
50\% of the inputs.
In Figure \ref{frac-cont} we show the fraction of recovered $B/T$
within different confidence levels. From that figure it is evident
that more than 90\% of the galaxies are well within 50\% of the input
value.

In Figure \ref{in-out8} it can be seen that the bulge parameters are
 underestimated for some cases. We found that this is largely due to the
overestimation of sky value by SEXTRACTOR. Due to this, if we fix the
background value at the \sex value while fitting, we obtain incorrect 
bulge parameters. The result can be improved by leaving the
background free (as we have done in our tests), 
but only a better algorithm to determine the sky can get rid of this
problem completely.

\subsection{Sensitivity to \sex input parameters}

While running Pymorph, the user has some flexibility in choosing input
parameters to SEXTRACTOR. It is important to test whether changes in \sex
input parameters affects the final results, when processing real
data. As a simple test, we used SDSS images of 160 galaxies in the $i$
band randomly selected (from within 4 morphological classes in equal
numbers) from the catalogue of $\sim14000$ visualy classified galaxies
from \nocite{nai10}{Nair} \& {Abraham} (2010). Our sample of 160 galaxies includes
nearly equal number of all the morphological types (ellipticals,
lenticulars, early type spirals, late type spirals).  Our galaxies
were selected to exclude those with bars/rings or any morphological
components, other than bulge and disk.

The input parameters that are likely to change galaxy structural
parameters significantly, are those related to the detection threshold
and background estimation. Therefore, we checked whether the final
output of \gal changes significantly as these parameters of \sex are
changed. We experimented with a detection threshold varying between
0.5 and 2.0.  We checked whether the output varies if one uses GLOBAL
background instead of LOCAL. We also varied the size of the background
square used in \sex (64 and 128 pixels wide). In all these
tests, there was no systematic deviation in extracted parameters, and
an overwhelming fraction of galaxies were consistently fit.

\section{Parallel PyMorph (PPyMorph)}
\label{parallel-sec}
When one thinks of the amount of data available from large
astronomical surveys today and volumes that will be obtained with
upcoming surveys, the need for parallisation of astronomical software,
whereever feasible, emerges naturally. For the estimation of
structural parameters, PyMorph needs significant computational time
when operating on large samples. We, therefore implemented PyMorph in
the parallel mode to make use of large number of CPUs and process
significant amount of data in a short time. The architecture of
PPyMorph is simple. It uses the Single Program, Multiple Data
technique. In this technique, we send different galaxy data to
different processors in a cluster to achieve coarse parallisation. Each
processor runs PyMorph on these data, and finally, all the results are
collected together. We use the Python \textit{pypar} module
extensively in PPyMorph.

\begin{figure*}
 \centering
 \includegraphics[scale=0.5]{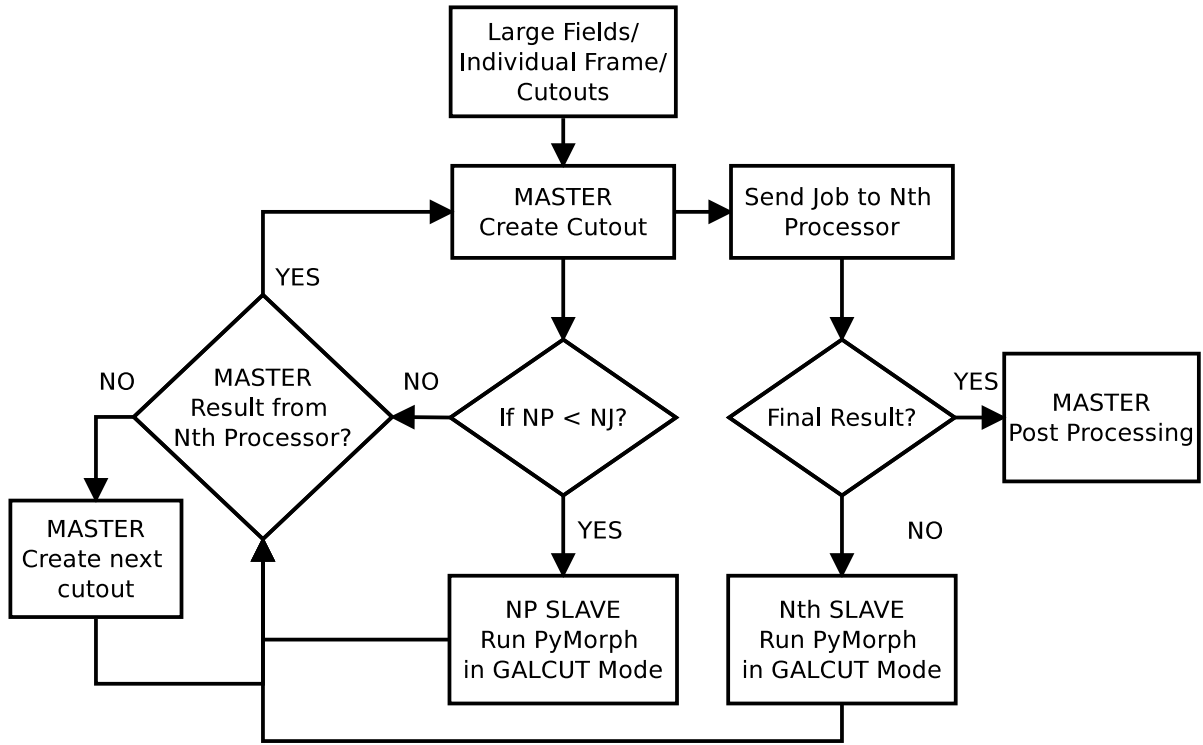}
 \caption{The architecture of Parallel PyMorph. NP and NJ are the number
of processors (SLAVES) and number of jobs (number of galaxies). Nth processor is
the first available SLAVE.}
 \label{parallel}
\end{figure*}

Figure \ref{parallel} shows the architecture of parallel PyMorph. Here
the user has the freedom to give input images in a variety of ways. It
is possible to give a large frame(s) which contains several objects or
cutouts of objects. In all the cases PPyMorph assigns one processor as
the MASTER and that creates cutouts of the galaxies. These are then
sent, one by one, to an available processor in the cluster which is
called a SLAVE. The SLAVE calls PyMorph to run on this cutout image in
\textit{galcut} mode. While the SLAVE is working on the given cutout,
the MASTER searches for an unoccupied SLAVE and assigns another galaxy to
it. This continues until all the available SLAVES are engaged. Then
MASTER readies cutout of additional galaxies in the list and waits, until one of
the SLAVES finishes processing the galaxy assigned to it. When
it finds a free SLAVE it fires the next job to that particular
SLAVE. This process continues until the MASTER has no more galaxies
left to fit. At this stage, the MASTER starts compilation of all the
results by the SLAVES and creates final outputs (plots, html, csv
etc.) for all galaxies.

In the non-parallel version, the pipeline takes $\sim 100$ sec
  to generate all the structural parameters  for a cutout
  of size $240 \times 240$ on a computer with a Intel Core2 Duo
  CPU at 1.5GHz with 2GB RAM. In the parallel version, as the number of
  processors increases 10 fold, the required time decreases $\sim
  6$ fold. 

\section{Summary}

We have presented a new software pipeline, PyMorph, to determine the
structural parameters of galaxies in an automated way. We have
described the methods implemented in the program. In the best cases,
PyMorph uses a relatively small number of user-specified parameters,
as compared to traditional fitting procedures. It makes extensive use
of \sex and GALFIT. PyMorph tries to obtain a global minimum from \gal
through clever use of the input parameters. We have tested the program
for simulated data and compared our results with earlier published
results. To increase processing speed, we have developed a parallel
form of the pipeline. Although PyMorph currently employs the popular
\gal software as the minimisation engine, it is flexible enough to
allow the user to replace \gal with some other galaxy structural
decomposition software, by modifying Pymorph.

The application of PyMorph ranges from individual images to large
surveys. We have ourselves used the Pymorph pipeline for a study of
galaxy morphology in clusters at moderate redshift
\nocite{vik10}({Vikram} {et~al.} 2010). Since it is implemented in \textit{Python} PyMorph is
largely OS independent. Also the implementation in Python make the
code reusable. PPyMorph makes it possible for astronomers to get structural
parameters of a large number of galaxies in a short time. Given the
complexity of the Pymorph package, we believe that some users will
prefer to use a web enabled interface to Pymorph to obtain structural
parameters for their galaxy images, without having to install the full
Pymorph package.  We are in the process of designing a Virtual
Observatory compatible web interface to PyMorph in collaboration with
VO-India.

A user manual of PyMorph is available at
https://www.iucaa.ernet.in/$\sim$vvinuv/UsersManual with detailed
discussions on the input and output parameters. PyMorph source code is available
on request.

\label{summ}

\section*{Acknowledgments}

We thank the anonymous referee for useful comments which have greatly
improved this paper. We thank  K. Indulekha
, Swara Ravindranath and  Sajeeth Ninan Philip for useful discussions.
Vinu Vikram acknowledges financial support from the
Council of Scientific and Industrial Research (CSIR). He also thanks IUCAA for 
the excellent hospitality provided during this work. 
%% \bibliography

\label{lastpage}


\begin{thebibliography}{}

\bibitem[{Abazajian}, {Adelman-McCarthy},  {Ag{\"u}eros}, {Allam}, {Allende Prieto}, {An}, {Anderson}, {Anderson},  {Annis}, {Bahcall}, {Bailer-Jones}, {Barentine}, {Bassett}, {Becker},  {Beers}, {Bell}, {Belokurov}, {Berlind}, {Berman}, {Bernardi}, {Bickerton},  {Bizyaev}, {Blakeslee}, {Blanton}, {Bochanski}, {Boroski}, {Brewington},  {Brinchmann}, {Brinkmann}, {Brunner}, {Budav{\'a}ri}, {Carey}, {Carliles},  {Carr}, {Castander}, {Cinabro}, {Connolly}, {Csabai}, {Cunha}, {Czarapata},  {Davenport}, {de Haas}, {Dilday}, {Doi}, {Eisenstein}, {Evans}, {Evans},  {Fan}, {Friedman}, {Frieman}, {Fukugita}, {G{\"a}nsicke}, {Gates},  {Gillespie}, {Gilmore}, {Gonzalez}, {Gonzalez}, {Grebel}, {Gunn},  {Gy{\"o}ry}, {Hall}, {Harding}, {Harris}, {Harvanek}, {Hawley}, {Hayes},  {Heckman}, {Hendry}, {Hennessy}, {Hindsley}, {Hoblitt}, {Hogan}, {Hogg},  {Holtzman}, {Hyde}, {Ichikawa}, {Ichikawa}, {Im}, {Ivezi{\'c}}, {Jester},  {Jiang}, {Johnson}, {Jorgensen}, {Juri{\'c}}, {Kent}, {Kessler}, {Kleinman},  {Knapp}, {Konishi}, {Kron}, {Krzesinski}, {Kuropatkin}, {Lampeitl},  {Lebedeva}, {Lee}, {Lee}, {Leger}, {L{\'e}pine}, {Li}, {Lima}, {Lin}, {Long},  {Loomis}, {Loveday}, {Lupton}, {Magnier}, {Malanushenko}, {Malanushenko},  {Mandelbaum}, {Margon}, {Marriner}, {Mart{\'{\i}}nez-Delgado}, {Matsubara},  {McGehee}, {McKay}, {Meiksin}, {Morrison}, {Mullally}, {Munn}, {Murphy},  {Nash}, {Nebot}, {Neilsen}, {Newberg}, {Newman}, {Nichol}, {Nicinski},  {Nieto-Santisteban}, {Nitta}, {Okamura}, {Oravetz}, {Ostriker}, {Owen},  {Padmanabhan}, {Pan}, {Park}, {Pauls}, {Peoples}, {Percival}, {Pier}, {Pope},  {Pourbaix}, {Price}, {Purger}, {Quinn}, {Raddick}, {Fiorentin}, {Richards},  {Richmond}, {Riess}, {Rix}, {Rockosi}, {Sako}, {Schlegel}, {Schneider},  {Scholz}, {Schreiber}, {Schwope}, {Seljak}, {Sesar}, {Sheldon}, {Shimasaku},  {Sibley}, {Simmons}, {Sivarani}, {Smith}, {Smith}, {Smol{\v c}i{\'c}},  {Snedden}, {Stebbins}, {Steinmetz}, {Stoughton}, {Strauss}, {Subba Rao},  {Suto}, {Szalay}, {Szapudi}, {Szkody}, {Tanaka}, {Tegmark}, {Teodoro},  {Thakar}, {Tremonti}, {Tucker}, {Uomoto}, {Vanden Berk}, {Vandenberg},  {Vidrih}, {Vogeley}, {Voges}, {Vogt}, {Wadadekar}, {Watters}, {Weinberg},  {West}, {White}, {Wilhite}, {Wonders}, {Yanny}, {Yocum}, {York}, {Zehavi},  {Zibetti}, \& {Zucker} 2009]{aba09}
{Abazajian} K.~N., {Adelman-McCarthy} J.~K., {Ag{\"u}eros} M.~A., {Allam}  S.~S., {Allende Prieto} C., {An} D., {Anderson} K.~S.~J., {Anderson} S.~F., {et al.}, 2009, \apjs, 182, 543

\bibitem[{Abraham}, {Tanvir}, {Santiago}, {Ellis},  {Glazebrook}, \& {van den Bergh} 1996]{abr96}
{Abraham} R.~G., {Tanvir} N.~R., {Santiago} B.~X., {Ellis} R.~S., {Glazebrook}  K., {van den Bergh} S., 1996, \mnras, 279, L47

\bibitem[{Barway}, {Kembhavi}, {Wadadekar},  {Ravikumar}, \& {Mayya} 2007]{bar07}
{Barway} S., {Kembhavi} A., {Wadadekar} Y., {Ravikumar} C.~D., {Mayya} Y.~D.,  2007, \apjl, 661, L37

\bibitem[{Barway}, {Wadadekar}, {Kembhavi}, \&  {Mayya} 2009]{bar09}
{Barway} S., {Wadadekar} Y., {Kembhavi} A.~K., {Mayya} Y.~D., 2009, \mnras,  394, 1991

\bibitem[{Bertin} \& {Arnouts} 1996]{ber96}
{Bertin} E., {Arnouts} S., 1996, \aaps, 117, 393

\bibitem[{Byun} \& {Freeman} 1995]{byu95}
{Byun} Y.~I., {Freeman} K.~C., 1995, \apj, 448, 563

\bibitem[{Conselice} 2003]{con03}
{Conselice} C.~J., 2003, \apjs, 147, 1

\bibitem[{de Jong} 1996]{jon96}
{de Jong} R.~S., 1996, \aaps, 118, 557

\bibitem[{de Souza}, {Gadotti}, \& {dos  Anjos} 2004]{sou04}
{de Souza} R.~E., {Gadotti} D.~A., {dos Anjos} S., 2004, \apjs, 153, 411

\bibitem[{Frei}, {Guhathakurta}, {Gunn}, \&  {Tyson} 1996]{fre96}
{Frei} Z., {Guhathakurta} P., {Gunn} J.~E., {Tyson} J.~A., 1996, \aj, 111, 174

\bibitem[{Gadotti} 2008]{gad08}
{Gadotti} D.~A., 2008, \mnras, 384, 420

\bibitem[{H{\"a}ussler}, {McIntosh}, {Barden},  {Bell}, {Rix}, {Borch}, {Beckwith}, {Caldwell}, {Heymans}, {Jahnke}, {Jogee},  {Koposov}, {Meisenheimer}, {S{\'a}nchez}, {Somerville}, {Wisotzki}, \&  {Wolf} 2007]{hau07}
{H{\"a}ussler} B., {McIntosh} D.~H., {Barden} M., {Bell} E.~F., {Rix} H.,  {Borch} A., {Beckwith} S.~V.~W., {Caldwell} J.~A.~R., {et al.}, 2007, \apjs, 172, 615

\bibitem[{James} 1994]{jam94}
{James} F., 1994, {MINUIT: Function Minimization and Error Analysis (CERN  Program Libr. Long Writeup D506) (version 94.1; Geneva: CERN)}

\bibitem[{Khosroshahi}, {Wadadekar}, \&  {Kembhavi} 2000]{kho00}
{Khosroshahi} H.~G., {Wadadekar} Y., {Kembhavi} A., 2000, \apj, 533, 162

\bibitem[{Laurikainen}, {Salo}, \& {Buta} 2005]{lau05}
{Laurikainen} E., {Salo} H., {Buta} R., 2005, \mnras, 362, 1319

\bibitem[{Lintott}, {Schawinski}, {Slosar}, {Land},  {Bamford}, {Thomas}, {Raddick}, {Nichol}, {Szalay}, {Andreescu}, {Murray}, \&  {Vandenberg} 2008]{lin08}
{Lintott} C.~J., {Schawinski} K., {Slosar} A., {Land} K., {Bamford} S.,  {Thomas} D., {Raddick} M.~J., {Nichol} R.~C., {et al.}, 2008, \mnras, 389, 1179

\bibitem[{Lotz}, {Primack}, \& {Madau} 2004]{lot04}
{Lotz} J.~M., {Primack} J., {Madau} P., 2004, \aj, 128, 163

\bibitem[{MacArthur}, {Courteau}, \&  {Holtzman} 2003]{mac03}
{MacArthur} L.~A., {Courteau} S., {Holtzman} J.~A., 2003, \apj, 582, 689

\bibitem[{Nair} \& {Abraham} 2010]{nai10}
{Nair} P.~B., {Abraham} R.~G., 2010, \apjs, 186, 427

\bibitem[{Peng}, {Ho}, {Impey}, \& {Rix} 2002]{pen02}
{Peng} C.~Y., {Ho} L.~C., {Impey} C.~D., {Rix} H., 2002, \aj, 124, 266

\bibitem[{Peng}, {Ho}, {Impey}, \& {Rix} 2010]{pen10}
---, 2010, \aj, 139, 2097

\bibitem[{Press}, {Teukolsky}, {Vetterling}, \&  {Flannery} 1992]{pre92}
{Press} W.~H., {Teukolsky} S.~A., {Vetterling} W.~T., {Flannery} B.~P., 1992,  {Numerical recipes in C. The art of scientific computing}, {Press, W.~H.,  Teukolsky, S.~A., Vetterling, W.~T., \& Flannery, B.~P. }, ed.

\bibitem[{Ravindranath}, {Ho}, {Peng},  {Filippenko}, \& {Sargent} 2001]{rav01}
{Ravindranath} S., {Ho} L.~C., {Peng} C.~Y., {Filippenko} A.~V., {Sargent}  W.~L.~W., 2001, \aj, 122, 653

\bibitem[{Rix}, {Barden}, {Beckwith}, {Bell}, {Borch},  {Caldwell}, {H{\"a}ussler}, {Jahnke}, {Jogee}, {McIntosh}, {Meisenheimer},  {Peng}, {Sanchez}, {Somerville}, {Wisotzki}, \& {Wolf} 2004]{rix04}
{Rix} H., {Barden} M., {Beckwith} S.~V.~W., {Bell} E.~F., {Borch} A.,  {Caldwell} J.~A.~R., {H{\"a}ussler} B., {Jahnke} K., {et al.}, 2004, \apjs, 152, 163

\bibitem[{Scoville}, {Aussel}, {Brusa}, {Capak},  {Carollo}, {Elvis}, {Giavalisco}, {Guzzo}, {Hasinger}, {Impey}, {Kneib},  {LeFevre}, {Lilly}, {Mobasher}, {Renzini}, {Rich}, {Sanders}, {Schinnerer},  {Schminovich}, {Shopbell}, {Taniguchi}, \& {Tyson} 2007]{sco07}
{Scoville} N., {Aussel} H., {Brusa} M., {Capak} P., {Carollo} C.~M., {Elvis}  M., {Giavalisco} M., {Guzzo} L., {et al.}, 2007, \apjs, 172, 1

\bibitem[{Sersic} 1968]{ser68}
{Sersic} J.~L., 1968, {Atlas de galaxias australes}, {Sersic, J.~L.}, ed.

\bibitem[{Simard} 1998]{sim98}
{Simard} L., 1998, in Astronomical Society of the Pacific Conference Series,  Vol. 145, Astronomical Data Analysis Software and Systems VII, {R.~Albrecht,  R.~N.~Hook, \& H.~A.~Bushouse}, ed., pp. 108--+

\bibitem[{Simard}, {Willmer}, {Vogt}, {Sarajedini},  {Phillips}, {Weiner}, {Koo}, {Im}, {Illingworth}, \& {Faber} 2002]{sim02}
{Simard} L., {Willmer} C.~N.~A., {Vogt} N.~P., {Sarajedini} V.~L., {Phillips}  A.~C., {Weiner} B.~J., {Koo} D.~C., {Im} M., {et al.}, 2002, \apjs, 142, 1

\bibitem[{Skrutskie}, {Cutri}, {Stiening},  {Weinberg}, {Schneider}, {Carpenter}, {Beichman}, {Capps}, {Chester},  {Elias}, {Huchra}, {Liebert}, {Lonsdale}, {Monet}, {Price}, {Seitzer},  {Jarrett}, {Kirkpatrick}, {Gizis}, {Howard}, {Evans}, {Fowler}, {Fullmer},  {Hurt}, {Light}, {Kopan}, {Marsh}, {McCallon}, {Tam}, {Van Dyk}, \&  {Wheelock} 2006]{skr06}
{Skrutskie} M.~F., {Cutri} R.~M., {Stiening} R., {Weinberg} M.~D., {Schneider}  S., {Carpenter} J.~M., {Beichman} C., {Capps} R., {et al.}, 2006,  \aj, 131, 1163

\bibitem[{Vikram}, {Wadadekar}, {Kembhavi}, \&  {Vijayagovindan} 2010]{vik10}
{Vikram} V., {Wadadekar} Y., {Kembhavi} A.~K., {Vijayagovindan} G.~V., 2010,  \mnras, 401, L39

\bibitem[{Wadadekar}, {Robbason}, \&  {Kembhavi} 1999]{wad99}
{Wadadekar} Y., {Robbason} B., {Kembhavi} A., 1999, \aj, 117, 1219

\bibitem[{York}, {Adelman}, {Anderson}, {Anderson},  {Annis}, {Bahcall}, {Bakken}, {Barkhouser}, {Bastian}, {Berman}, {Boroski},  {Bracker}, {Briegel}, {Briggs}, {Brinkmann}, {Brunner}, {Burles}, {Carey},  {Carr}, {Castander}, {Chen}, {Colestock}, {Connolly}, {Crocker}, {Csabai},  {Czarapata}, {Davis}, {Doi}, {Dombeck}, {Eisenstein}, {Ellman}, {Elms},  {Evans}, {Fan}, {Federwitz}, {Fiscelli}, {Friedman}, {Frieman}, {Fukugita},  {Gillespie}, {Gunn}, {Gurbani}, {de Haas}, {Haldeman}, {Harris}, {Hayes},  {Heckman}, {Hennessy}, {Hindsley}, {Holm}, {Holmgren}, {Huang}, {Hull},  {Husby}, {Ichikawa}, {Ichikawa}, {Ivezi{\'c}}, {Kent}, {Kim}, {Kinney},  {Klaene}, {Kleinman}, {Kleinman}, {Knapp}, {Korienek}, {Kron}, {Kunszt},  {Lamb}, {Lee}, {Leger}, {Limmongkol}, {Lindenmeyer}, {Long}, {Loomis},  {Loveday}, {Lucinio}, {Lupton}, {MacKinnon}, {Mannery}, {Mantsch}, {Margon},  {McGehee}, {McKay}, {Meiksin}, {Merelli}, {Monet}, {Munn}, {Narayanan},  {Nash}, {Neilsen}, {Neswold}, {Newberg}, {Nichol}, {Nicinski}, {Nonino},  {Okada}, {Okamura}, {Ostriker}, {Owen}, {Pauls}, {Peoples}, {Peterson},  {Petravick}, {Pier}, {Pope}, {Pordes}, {Prosapio}, {Rechenmacher}, {Quinn},  {Richards}, {Richmond}, {Rivetta}, {Rockosi}, {Ruthmansdorfer}, {Sandford},  {Schlegel}, {Schneider}, {Sekiguchi}, {Sergey}, {Shimasaku}, {Siegmund},  {Smee}, {Smith}, {Snedden}, {Stone}, {Stoughton}, {Strauss}, {Stubbs},  {SubbaRao}, {Szalay}, {Szapudi}, {Szokoly}, {Thakar}, {Tremonti}, {Tucker},  {Uomoto}, {Vanden Berk}, {Vogeley}, {Waddell}, {Wang}, {Watanabe},  {Weinberg}, {Yanny}, \& {Yasuda} 2000]{yor00}
{York} D.~G., {Adelman} J., {Anderson} Jr. J.~E., {Anderson} S.~F., {Annis} J.,  {Bahcall} N.~A., {Bakken} J.~A., {Barkhouser} R., {et al.}, 2000, \aj, 120, 1579

\end{thebibliography}
\end{document}